\documentclass{an_art}
\usepackage{graphicx}

\newcommand{\gtrsim}{\mathrel{\hbox{\rlap{\lower.55ex \hbox {$\sim$}}
                   \kern-.3em \raise.4ex \hbox{$>$}}}}
\newcommand{\lesssim}{\mathrel{\hbox{\rlap{\lower.55ex \hbox {$\sim$}}
                   \kern-.3em \raise.4ex \hbox{$<$}}}}

\begin{document}

\begin{titlepage}

\setcounter{page}{000}
\headnote{Astron.~Nachr.~000 (0000) 0, 000--000}
\makeheadline

\title {A `superoutburst' in XTE\,J1118+480}

\author{
{\sc Erik~Kuulkers\footnote{Also: Astronomical Institute, Utrecht University,
P.O.\ Box 80000, 3507 TA Utrecht, The Netherlands}} , Utrecht, The Netherlands \\
\medskip
{\small Space Research Organization Netherlands}
}
\date{Received; accepted} 
\maketitle

\summary
I propose that the properties of the two outbursts observed in the X-ray transient
XTE\,J1118+480 in 2000 are akin to superoutbursts of SU\,UMa stars.
In these systems a `normal' outburst immediately precedes a 5--10 times longer (`super')outburst.
The optical light curve of the outbursts of XTE\,J1118+480 is remarkably similar to that
seen in some SU\,UMa stars, such as UV\,Per and T\,Leo, where the precursor outburst is distinct from
the superoutburst, but the time scales are a factor of $\sim$15 different.
The first outburst of XTE\,J1118+480 was relatively short ($\sim$1~month) while the second outburst
was $\sim$5 times longer. During the second outburst superhumps were seen, a feature characteristic
for superoutbursts. The gap of about a month between the two outbursts is longer in X-rays with
respect to the optical, a feature not previously recognized for X-ray transients. Also in SU\,UMa stars
the precursor outburst becomes more distinct at shorter wavelengths.
Finally, I show that the time of appearance of the superhumps in XTE\,J1118+480 is consistent with
the expected superhump growth time, if the superhump mechanism was triggered during the first outburst.
I conclude that the similarity in outburst behaviour in the two types of systems
provides further support that a common mechanism is at work to start the long (`super')outbursts.
END

\keyw
binaries: close --- stars: individual (XTE\,J1118+480, UV\,Per, T\,Leo) --- novae, cataclysmic variables --- X-rays
END

\end{titlepage}

%
\section{Introduction}

Outbursts in dwarf novae, a subgroup of the cataclysmic variables 
(CVs)\footnote{Binary stars in which a Roche-lobe filling 
donor star transfers matter towards a white dwarf through an (unstable) accretion disk, 
see Warner 1995a for a review.}, and X-ray novae or (soft) X-ray transients (SXTs), 
a subgroup of the low-mass X-ray binaries\footnote{In these binaries 
matter is transferred to a neutron star or black hole (see Tanaka \&\ Lewin 1995; Tanaka \&\ Shibazaki 1996,
for reviews).}, are thought to be due to an enhancement of the mass transfer rate through the
accretion disk (see, e.g., Lasota 1996). 

A subgroup of the dwarf novae, the SU\,UMa stars (see Warner 1995a, 1995b for reviews) shows both
short and long outbursts. These outbursts are referred to as `normal' outbursts 
and `super'-outbursts, respectively. Normal outbursts have durations on the 
order of days to a week, whereas superoutbursts may last for a week to several 
weeks. Only during superoutbursts a periodic modulation appears,
with a period which is a few percent longer than the orbital period: the `super'-hump.
The superoutburst recurrence times are from weeks to months to decades.
In between superoutbursts one or more normal outbursts may occur.
The longer superoutburst is most probably the consequence
of a (temporal) increase in the transfer rate from the donor, possibly 
triggered by heating of the companion (e.g., Smak 1995, 1996).

The superhumps are thought to arise due to the excitation of the 3:1
orbital resonance in the disk (Whitehurst 1988; Lubow 1991), causing the disk to become eccentric and
to precess, and may be triggered only in
systems with ratios of the donor mass to the compact mass, $q$,
of less then $\sim$0.33 (e.g., Whitehurst \&\ King 1991; Whitehurst 1994).
This requirement is reached in the SU\,UMa stars.
Since in most of the SXTs the compact object is a black hole (e.g., Charles 1998),
they have generally small mass ratios as well, and their disks are therefore also subject to the
3:1 resonance (e.g., Whitehurst 1994).
Superhumps have indeed been reported during outbursts of a few
SXTs (see O'Donoghue \&\ Charles, 1996, and references therein).
The tidal stresses in the outer disk cause
a local enhancement in the intensity, resulting in a hump in the light curve
when the enhancement appears in our line of sight in the SU\,UMa stars,
with a period a few per cent longer than the orbital period.
In the SXTs, however, the optical light is dominated by the X-ray reprocessing,
and there superhumps can be seen because of the varying area of the eccentric disk
(Haswell et al.\ 2001).

\begin{figure}
\resizebox{115mm}{!}{\includegraphics[angle=-90, origin=c, clip, bb=78 77 512 588]{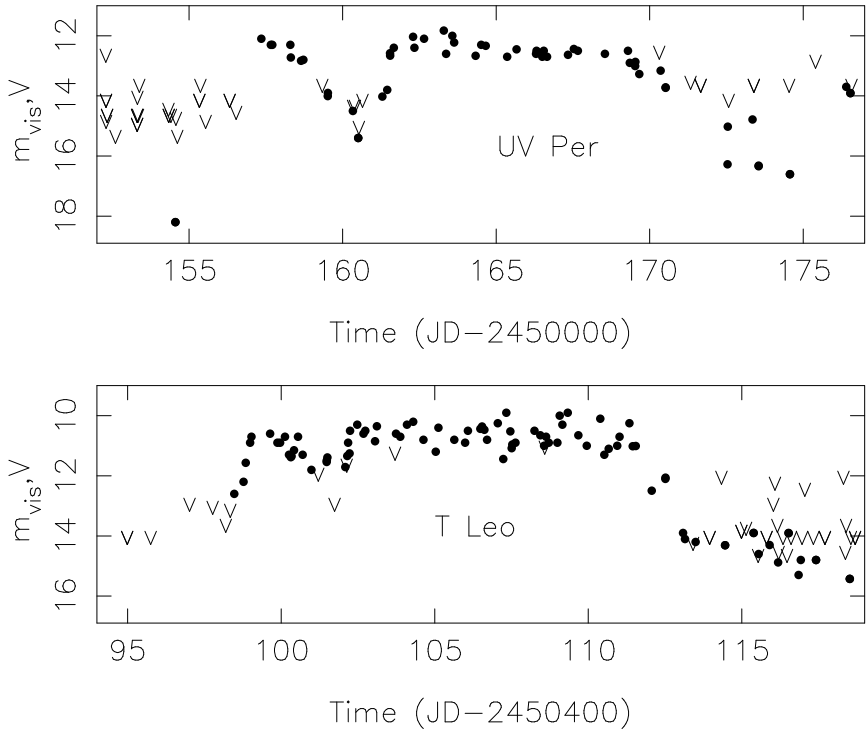}}
\vspace{-1.0cm}
\hfill
\parbox[b]{55mm}{
\caption{Optical light curves of the superoutbursts of UV\,Per (upper panel) in 1996 
and T\,Leo in 1997 (lower panel; see also Howell et al.\ 1999). Optical data comprise 
visual estimates 
and CCD measurements reported to \mbox{VSNET} (Variable Star NETwork), the AAVSO, the VSOLJ and the AFOEV. 
For UV\,Per I also
used data from the BAAVSS/TA. Upper limits
are denoted as $\vee$. The data have rebinned to a time resolution of 0.1\,day. The
quiescent levels of these systems are 17.1--17.9 and 15.2--15.9~mag, respectively
(see Ritter \&\ Kolb 1998).
Note the drop in magnitude in the light curve a few days after the start of the outburst,
and the flat top during the main part of the outburst. UV\,Per also shows rebrightenings
after the main superoutburst.}
\vspace{1.0cm}
\label{uvper_tleo}}
\end{figure}

The superoutbursts in SU\,UMa stars are immediately preceded by a normal
outburst (Bateson 1977; Marino \&\ Walker 1979; van der Woerd \&\
van Paradijs 1987). Sometimes the precursor normal outburst is well separated from
the superoutburst (e.g., T\,Leo [see also Kato 1997] 
and UV\,Per: see Fig.~1); at other times this separation is not clear, i.e.\ the normal
outburst profile smoothly changes into the superoutburst profile (Marino \&\ Walker 1979).
In the latter case at least the rise to maximum brightness is similar to that in 
normal outbursts (see e.g., Warner 1995b, and references therein).
At shorter wavelengths, however, the separation becomes more evident.
A nice example is the 1984 superoutburst of VW\,Hyi
(Pringle et al.\ 1987; Polidan \&\ Holberg 1987): the {\it Voyager}
far-UV outburst light curve showed a clear precursor akin to a normal outburst,
while in the optical this was not evident. Another example is the 1987 superoutburst
of Z\,Cha: the precursor was not obvious in the V-band, while it became progressively more distinct
in the near-UV wavelengths (Kuulkers et al.\ 1991).

The outbursts of SXTs last typically much longer ($\sim$months) and are more
energetic, basically due to the deeper potential well of
the compact object and the presence of irradiation of the disk by X-rays 
(see, e.g, van Paradijs \&\ McClintock 1994, 1995). 
Also, their recurrence times are generally much longer than those in dwarf novae, i.e.\ $\sim$years to decades. 
As the number of known X-ray transients began to grow
(e.g., White, Kaluzienski \&\ Swank 1984; see also Chen, Shrader \&\ Livio 1997),
it was already soon realized that outbursts in SXTs may be triggered by the same mechanism
as those in dwarf novae (van Paradijs \&\ Verbunt 1984).
In fact, it is an extreme group of the SU\,UMa stars (i.e\ those
having superoutbursts with very long recurrence times, $\sim$years to decades, 
but [almost] no normal outburst), which shows
the best resemblance with the SXTs (Lasota 1996; Kuul\-kers, Howell \&\ van Paradijs 1996; Kuul\-kers 1999, 
2000; Hellier 2001).
The members of this group are referred to as WZ\,Sge stars (e.g., Bailey 1979), or 
sometimes as `tremendous outburst amplitude dwarf novae' (TOADs, Howell, Szkody \&\ Cannizzo 1995).

Some of the superoutbursts of WZ\,Sge stars have been seen to exhibit a precursor
as well (see, e.g., Fig.~1). However, it is not clear whether {\it all} 
the superoutbursts in WZ\,Sge stars (such as those in WZ\,Sge itself) are preceded by a normal outburst
(but see the Discussion).
X-ray precursors to the main long X-ray outbursts have been seen in several SXTs. 
Sometimes, even a sequence of outbursts appear after a long quiescent time (see Chen et al.\ 1997, and 
references therein).
Since SXT outbursts are always first detected in X-rays (except for those with more 
frequent recurrent outbursts, like Aql\,X-1 and GRO\,J1655$-$40), the optical identification
almost always comes later. It was therefore hitherto not possible to judge whether precursors also exist
in the optical (or other wavelengths). For the few more frequent outbursting SXTs, when both optical 
and X-ray observations are available, either the coverage is rather sparse, or it is not clear
whether a wavelength dependent precursor exist. Note, however, that at least the start of the outburst
of these SXTs shows similar features to that seen in the normal outbursts
of SU\,UMa stars, i.e.\ a delay in the start in the X-ray outburst with respect to the optical
(GRO\,J1655$-$40: Orosz et al.\ 1997; Aql\,X-1: Shahbaz et al.\ 1998; XTE\,J1550$-$564: 
compare Smith et al.\ 2000 with Jain \&\ Bailyn 2000). In SU\,UMa stars this is seen in the 
form of a UV-delay with respect to the optical of up to half a day 
(Warner 1995a, and references therein). This suggests that (most of) the SXT outbursts start 
as a `normal' outburst (see, e.g., Hameury et al.\ 1997, in the case of GRO\,J1655$-$40), 
triggers a (temporal) increase in the transfer rate from the 
secondary by X-ray heating (e.g., Esin, Lasota \&\ Hynes 2000), and proceeds as a long (super)outburst.

\begin{figure}
\resizebox{115mm}{!}{\includegraphics[angle=-90, origin=c, clip, bb=78 83 418 588]{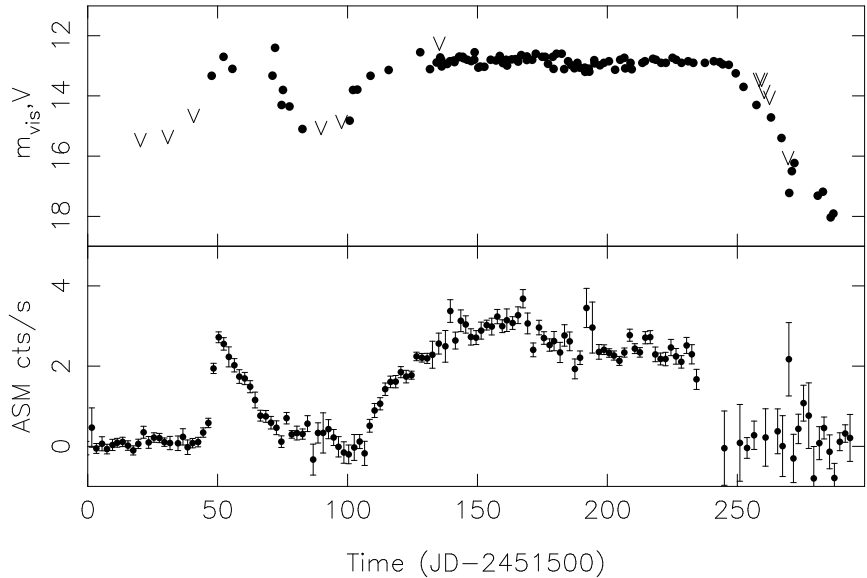}}
\vspace{-1.9cm}
\hfill
\parbox[b]{55mm}{
\caption{Optical (upper panel) and X-ray (lower panel) light curve of the two
outbursts seen so far of XTE\,J1118+480. 
Optical data comprise visual estimates, 
photographic and CCD measurements reported to \mbox{VSNET},
see also Uemura et al.\ (2000c). Upper limits
are denoted as $\vee$. In quiescence the source reaches $\sim$19~mag (Uemura et al.\ 2000a, 2000c).
The X-ray data are from the All Sky Monitor (2--12\,keV) onboard the 
{\it Rossi X-ray Timing Explorer}. All data have been rebinned to a time resolution of
2~days.}
\vspace{1.9cm}
\label{xtej1118}}
\end{figure}

Outbursts of both SU\,UMa stars and SXTs enables us study unstable accretion disks, as well as the
occurrence of tidal resonances in these disks. Because of the number of SU\,UMa stars known 
and the higher frequency of occurrence of superoutbursts, observations of these stars may provide
us clues in understanding the behaviour of SXTs. On the other hand, the outbursts of SXTs last 
much longer, so processes can be studied in more detail, which may be rather shortlived in
SU\,UMa stars. Moreover, by comparison of the two classes, the effect of especially X-ray irradiation 
can be investigated.
The outbursts in a recently detected SXT, XTE\,J1118+480 had ample coverage 
in both X-rays and optical (e.g., Uemura et al.\ 2000c). In this paper, I speculate for the first time 
on the existence of a (wavelength dependent) precursor in an SXT outburst; I also
provide a possible interpretation of the observed characteristics of the superhumps found in this
system.

\section{XTE\,J1118+480 in outburst}

A high galactic latitude X-ray transient, XTE\,1118+480, was discovered on 2000 March 29 
in data from the All Sky Monitor (ASM) onboard the {\it Rossi X-ray Timing Explorer} 
(Remillard et al.\ 2000). Retrospective analysis showed that the X-ray 
outburst had already begun near March 5, and that another outburst had
occurred on 2000 January 2--29. The optical counterpart was reported soon after the 
X-ray discovery (Uemura, Kato \&\ Yamaoka 2000a, Uemura et al.\ 2000c). 
Archival photographic data (Uemura et al.\ 2000a, 2000c), 
as well as data from the {\it Robotic Optical Transient Search Experiment} (ROTSE; 
Wren \&\ McKay 2000) confirmed the earlier outburst
in January, as well as the rise to the second outburst, in the optical. 
XTE\,J1118+480 was found to be also active in $\gamma$-rays (Wilson \&\ McCollough 2000) 
and at radio wavelengths (Pooley \&\ Waldram 2000).
The source was in the so-called low/hard state commonly
seen in SXTs containing a black hole, at least after the discovery date 
(Revnivtsev, Gilfanov \&\ Churazov 2000, Hynes et al.\ 2000, Wood et al.\ 2000).
The characteristics of the second outburst of XTE\,J1118+480 are similar to that
seen during the `mini-outbursts' (or `reflares') in the SXT GRO\,J0422+32 
(both events featured low X-ray to optical luminosities, superhumps [see below], 
similar optical spectra, low/hard state, similar orbital period; Hynes et al.\ 2000, 
Dubus et al.\ 2001a).
Based on the characteristics described above, the compact object in XTE\,J1118+480 was 
believed to be a black hole. The confirmation came recently, when the dynamical evidence for 
the presence of a black hole was announced by McClintock et al.\ (2000, 2001) and Wagner et al.\ (2000):
f(M)$\sim$6\,M$_{\odot}$.

The mass ratio, $q$, is plausibly between 0.02--0.1\footnote{This
is consistent with the secondary star being close to a M0V star (Wagner et al.\ 2000).
A single M0V star has a mass of $\sim$0.5\,M$_{\odot}$ (Allen 1973); together with
the fact that the compact object must be more massive than 6\,M$_{\odot}$,
it follows that $q\lesssim 0.08$.} (Dubus et al.\ 2001a)
and one therefore expects to see superhumps. Indeed photometric modulations
were found soon after the discovery of the optical counterpart
at a period of $\sim$4.1\,hrs (Cook et al.\ 2000, Uemura et al.\ 2000b). 
Their period changed from 4.102 to 4.087\,hrs, while their shape changed from sinusoidal 
to `asymmetric hump' like (Uemura et al.\ 2000b). The latter bear resemblance to
superhumps seen in SU\,UMa stars; however, it was not clear whether the superhump evolved from
an orbital hump due to X-ray irradiation of the secondary (Uemura et al.\ 2000c), or that
both were in fact superhumps in a different fashion. 
In SU\,UMa stars and SXTs, the superhump period is always {\it longer} than the orbital period (although
CVs exist with superhump periods {\it less} than the orbital period; however,
none of them are SU\,UMa stars, see Patterson 1999). 
Moreover, the orbital period has been established recently: 4.08\,hr 
(McClintock et al.\ 2000, 2001, Wagner et al.\ 2000). 
Also, a (small) decay in the superhump period is generally seen during a superoutburst.
This, therefore, suggests that all of the modulations observed during the outburst
were superhumps. 

In Fig.~2 I show the optical and X-ray light curves of XTE\,J1118+480 during the two outbursts. 
It was already noted by Uemura et al.\ (2000c) that the first outburst 
lasted longer in the optical than X-rays. Moreover, they found that the 
second outburst started $\sim$10~days earlier in the optical with respect 
to the X-rays. As noted in the Introduction, delays between the X-ray and optical start of the outburst
are not uncommon in SXTs. A delay between the optical and X-rays can be
present during the rise to the first outburst as well, however, the 
optical data are too sparse during that period.
As can be seen in Fig.~2, the second outburst also started to decay earlier in 
the X-rays, with respect to the optical. This is also seen
in other SXT outbursts (e.g., A0620$-$00: see Kuul\-kers 1998), and is typical of 
X-ray irradiated disks (Dubus, Hameury \&\ Lasota 2001b).

The second outburst of XTE\,J1118+480 has a flat top shape in both the optical
and X-rays. Such flat tops are also not uncommon in SXT outbursts (see Chen et al.\ 1997).
Similar flat tops in the optical can be seen in outbursts of SU\,UMa stars
The resemblance of the outburst light curve of XTE\,J1118+480 with those of, for example, 
T\,Leo and UV\,Per (see Fig.~1), is rather striking. The only main difference is the 
difference in timescale, a factor of $\sim$15; this is similar to that seen 
when comparing other SXT and SU\,UMa stars (e.g., Kuulkers et al.\ 1996).

\section{Discussion}

Uemura et al.\ (2000c) concluded that the first and second outburst are reminiscent of an
`inside-out' and an `outside-in' type of outburst, respectively, and that the two types
are related to the two types of outbursts seen in 
the SU\,UMa stars, i.e.\ normal and superoutbursts.
I, however, go one step further and speculate that the two outbursts are in fact connected, 
i.e.\ the second outburst evolved from the first outburst.
This is similar to that seen in SU\,UMa stars, where a normal outburst precedes a superoutburst.
In XTE\,J1118+480 the precursor `normal' outburst is clearly distinguishable from
the second long (`super') outburst, resembling the T\,Leo and UV\,Per 
superoutburst light curves. Also the ratio of the durations of the two outbursts
are similar to the ratio of durations in the precursor normal outburst and superoutburst
in SU\,UMa stars (see Warner 1995b): $\sim$5. 
Moreover, the time between the maximum of the first outburst and the start 
of the second outburst of XTE\,J1118+480 is longer in X-rays with respect to the optical.
This is analogous to that seen in superoutbursts, where the precursor is more 
distinct at shorter wavelengths.
Between the precursor and the actual superoutburst in SU\,UMa stars
quiescence is never reached in the optical. Although this may be suggestive for XTE\,J1118+480
too (Fig.~2), there are only upper limits ($\sim$15~mag) for a $\sim$20~day interval.
Note that the decline from the precursor normal outburst in VW\,Hyi 
occurred more or less at the same time in the optical and far-UV (as is the case
for all normal outbursts in dwarf novae, see Warner 1995a) in contrast with that seen in XTE\,J1118+480. 
This is due to the absence of disk irradiation in SU\,UMa stars
(see Dubus et al.\ 2001b).

In order for the 3:1 orbital resonance to be excited to start the superhumps, 
the disk has to grow beyond its tidal radius. The precursor normal outburst may have 
provided the trigger for this, although the exact mechanism is still not clear 
(see, e.g., Osaki 1996; Smak 1996; Hameury, Lasota \&\ Warner 2000, for more details and discussions).
The time between the maximum of the precursor outburst and the
time of start of superhumps ranges from 1--3~days in normal SU\,UMa stars to up to 10\,days in 
WZ\,Sge stars (see Vogt 1993; Warner 1995a, 1995b). 
The superhumps in SU UMa stars start near the rise to the actual superoutburst maximum
(see, e.g., Warner 1995a), and therefore the timescale 
is effectively the time between precursor maximum and the time of start of the
actual superoutburst. In black hole SXTs and some WZ\,Sge stars 
the start of the superhumps is accompanied by a (brief) 
increase in luminosity, causing a secondary maximum on the way of decline.
(see, e.g., Kato et al.\ 1995; Kuulkers et al.\ 1996; Cannizzo 1998; Patterson et al.\ 1998; 
O'Donoghue 2000, and references therein). 
The time between the maximum of the outburst and the 
start of the superhumps in black hole SXTs is 45--55\,days (see, e.g., 
Charles \&\ O'Donoghue 1996; Kuul\-kers et al.\ 1996; Cannizzo 1998).
It may be that the part of the outburst before the secondary maximum in these systems
is the `precursor' to the part of the outburst after the secondary maximum. 
Interestingly, the secondary maximum becomes more pronounced in X-rays with respect to optical
in, e.g., GRO\,J0422+32, which may be related to a wavelength dependent precursor.
However, for other SXTs this is not so clear (see Kuulkers 1998, and 
references therein).

The time between the precursor maximum and the start of superhumps (or, equivalently, time of rise 
to superoutburst maximum or secondary maximum) is thought to be related
to the superhump growth time (see, e.g., Warner 1995b).
The superhump growth time is (ideally) proportional to 
P$_\mathrm{orb}$\,$q^{-2}$ (where P$_\mathrm{orb}$ is the orbital period; Lubow 1991; 
see also Warner 1995b), which is in excellent
agreement with that observed in SU\,UMa stars (one also has to take into account here 
whether the donor is a main sequence star or degenerate star, Warner 1995b).
For the black hole SXTs in which superhumps are seen, with P$_\mathrm{orb}\simeq 5$--10\,hr
and $q\simeq 0.04$--0.1, the superhump growth time 
is a factor of $\sim$30 larger than in SU\,UMa stars 
(as already recognized earlier, by, e.g., Ichikawa, Mineshige \&\ Kato 1994), indeed 
in the range observed. Now, for XTE\,J1118+480, with P$_\mathrm{orb}=4.08$\,hr
and $q\simeq 0.02-0.1$, this leads to a (rather large) range in the estimated growth times 
between 17--430~days. The time between the maximum of the first outburst 
and the start of the second outburst is on the order of 30--50\,days in XTE\,J1118+480.
If, as I speculate here, the first (`normal') outburst of XTE\,J1118+480 indeed 
evolved into the second outburst, this timescale may reflect the superhump growth time,
which is consistent with the estimated range in growth time.
Conversely, one might get a crude estimate
of $q$, assuming the observed superhump growth time is $\sim$30--50\,days, i.e.\ $q\simeq 0.06$--0.075;
this is also consistent with the value of $q$ inferred from the optical spectroscopic 
measurements of the secondary star and the range postulated by 
Dubus et al.\ (2001a; see Sect.\ 2).

If the superhumps in XTE\,J1118+480 developed during the second outburst,
no superhumps are expected during the first outburst in XTE\,J1118+480. 
Note that in SU\,UMa superoutbursts brightness modulations 
with periods equal to the orbital period have been seen before the start of superhumps
(see, e.g., Warner 1995a). Unfortunately, no observations exist to confirm whether there are 
any kind of modulations during the first outburst of XTE\,J1118+480.

Since the inclination at which one views XTE\,J1118+480 is likely to be low
and its distance is only $\sim$800\,pc (Hynes et al.\ 2000; Dubus et al.\ 2001a), 
the intrinsic X-ray luminosity from the inner parts of the accretion disk is low, and thus
X-ray irradiation may not be as strong as in other SXTs. As discussed in Sect.~2, the superhump
shape changed from sinusoidal to `asymmetric hump' like. The former are expected
if they are due to a periodically varying area of an eccentric (precessing) 
X-ray irradiated accretion disk (Haswell et al.\ 2001), while the latter are typical of SU\,UMa type of 
superhumps (see, e.g., Warner 1995a), i.e.\ due to tidal stresses in a non (or weak) X-ray 
irradiated accretion disk. This may suggest that irradiation dominated the optical 
light during the part of the second outburst when the superhumps had a sinusoidal shape, 
but when the outburst progressed irradiation became less strong, and the (outer) disk itself became the 
dominant source of optical light. 

I conclude that from a phenomenological comparison of the outburst characteristics of 
XTE\,J1118+480 with those seen in SU\,UMa stars, the two outbursts witnessed in XTE\,J1118+480
may be a superoutburst which evolved from a precursor normal outburst. This is the first indication 
of the presence of such a feature in SXTs. Close monitoring at various wavelengths 
of future outbursts of this system and of other SXTs may provide more examples of the
existence of precursors, and may help to gain more insight in the mechanism for
producing long ('super')outbursts and the occurrence of superhumps.
I end by noting that not all the SXTs have simple light curves
similar to that outlined here. Outbursts of, e.g., GRO\,J1655$-$40 (see Chen et al.\ 1997;
Hynes et al.\ 1998; Sobczak et al.\ 1999) and XTE\,J1550$-$564 (e.g., Jain et al.\ 1999)
are certainly more complex, and it remains to be seen if a similar reasoning, as I
have done for XTE\,J1118+480 applies to these systems as well.

\acknowledgements
I thank Jean in 't Zand, Jean-Pierre Lasota and Jerry Orosz for discussions,
and the referee for his valuable criticism.
I acknowledge the use of daily averaged quick-look results provided by the ASM/RXTE team.
In this research, I have also used, and acknowledge with thanks, data from the AAVSO International Database,
based on observations submitted to the AAVSO by variable star observers worldwide;
I also made use of the AFOEV (Association Francaise des Observateurs d'Etoiles
Variables) database, operated at CDS, France,
as well as of observations obtained by the VSOLJ (Variable Star Observers League in Japan),
the BAAVSS (British Astronomical Association Variable Star Section) and 'The Astronomer' 
organisation (TA), and observations reported to the International Mailing List on 
Variable Stars (\mbox{VSNET}\footnote{http://www.kusastro.kyoto-u.ac.jp/vsnet/}).
Part of this work is entirely attributed to ``the zeal and assiduity, not only
of the public observers, but also of the many private individuals, who nobly
sacrifice a great portion of their time and fortune to the laudable pursuit"
of these variables (Moyes 1831).



%
\addresses
\rf{Erik Kuulkers, 
Space Research Organization Netherlands, 
Sorbonnelaan 2, 
3584 CA Utrecht, 
The Netherlands,
e-mail: E.Kuulkers@sron.nl}

\end{document}